\documentclass[runningheads]{llncs}
\usepackage{multirow}
\usepackage[table,xcdraw]{xcolor}
\usepackage[T1]{fontenc}
\usepackage{graphicx}
\usepackage{hyperref}
\usepackage{color}

\begin{document}
\title{Controlling the Misinformation Diffusion in Social Media by the Effect of Different Classes of Agents}
\titlerunning{Controlling the misinformation in social media}
%
\author{Ali Khodabandeh Yalabadi\inst{1} \and Mehdi Yazdani-Jahromi\inst{2} \and Sina Abdidizaji\inst{1} \and Ivan Garibay\inst{1} \and Ozlem Ozmen Garibay\inst{1, 2}}
\authorrunning{A. Khodabandeh Yalabadi et al.}
%
\institute{Industrial Engineering and Management Systems, University of Central Florida, Orlando FL 32816, USA \and
Computer Science, University of Central Florida, Orlando FL 32816, USA\\
\email{\{yalabadi, yazdani, sina.abdidizaji, igaribay, ozlem\}@ucf.edu}}

%
\maketitle              
\begin{abstract}
The rapid and widespread dissemination of misinformation through social networks is a growing concern in today's digital age. This study focused on modeling fake news diffusion, discovering the spreading dynamics, and designing control strategies. A common approach for modeling the misinformation dynamics is SIR-based models. Our approach is an extension of a model called 'SBFC' which is a SIR-based model. This model has three states, Susceptible, Believer, and Fact-Checker. The dynamics and transition between states are based on neighbors' beliefs, hoax credibility, spreading rate, probability of verifying the news, and probability of forgetting the current state. Our contribution is to push this model to real social networks by considering different classes of agents with their characteristics. We proposed two main strategies for confronting misinformation diffusion. First, we can educate a minor class, like scholars or influencers, to improve their ability to verify the news or remember their state longer. The second strategy is adding fact-checker bots to the network to spread the facts and influence their neighbors' states. Our result shows that both of these approaches can effectively control the misinformation spread.

\keywords{misinformation  \and agent-based model \and social media \and fact-checking.}
\end{abstract}
\section{Introduction}
The rapid and widespread dissemination of misinformation through social networks is a growing concern in today's digital age. Misinformation can lead to harmful consequences, such as spreading false health information, polarization of public opinion, and even disrupting democratic processes. Understanding the dynamics of how misinformation spreads through social networks is crucial for developing effective strategies to mitigate its negative impacts.
Agent-based modeling (ABM) has emerged as a powerful tool for studying the spread of misinformation in social networks. ABM allows us to simulate individuals' behavior and interactions within a social network, making it an ideal method for analyzing the complex dynamics of information diffusion. In this paper, we build on the work of previous researchers to develop an ABM framework for studying the spread of misinformation in social networks.
Our approach draws on the findings of Tambuscio et al. \cite{tambuscio2015fact} and Sulis and Tambuscio \cite{sulis2020simulation}, who developed a model to simulate the spread of misinformation in social networks. Their work highlights the role of fact-checking interventions and network structure in mitigating the spread of misinformation. This SIR-based (Susceptible, Infected, and Recovered) model is a perfect model for testing our strategies because it can depict a clear confrontation between agents who believe the false news and agents who know the facts and are fighting against misinformation spread. We build on these insights by incorporating additional factors, such as individual characteristics and the influence of opinion leaders, in our ABM framework. So, we assume that not all the people (agents) in the network come from the same class with the same attributes and behaviors. Beskow and Carley \cite{beskow2019agent} introduced two critical agents of social networks: bots and influencers. These classes can heavily affect the dynamics of information spreading.

As an assumption, we categorized agents into four classes. 1) scholars who have a clearer understanding of information can verify the news better, remember the news credibility longer, and their community is concentrated in a cluster so they get more effect from each other as one cluster. 2) The influencers who have more influence because of their big neighborhood size, stay in their state (believer, fact-checker) longer, and could have a more verifying probability. 3) The normal agents who are the majority of the network, forget their state more frequently and have less verifying probability. 4) The bots or Super-spreaders who do not change their state and do not forget. They just act as spreaders in the network. They can be believer bots that are programmed to spread the hoax or they can be fact-checker bots that are programmed to mitigate the misinformation diffusion.
By modeling these assumptions and implementing the simulation, our ultimate goal is to use our ABM framework to gain insights into the factors that influence the spread of misinformation in social networks and to explore the effectiveness of different intervention strategies.
This paper is organized as follows. Section 2 will explain the previous literature related to our work. In section 3, method and implementation details have been discussed. We explored the results in section 4, and last but not least, we concluded our findings in section 5.

\section{Related Work}
Social media platforms are places where everybody has the ability to spread their words without a serious intervention of third-party filtering, fact-checking applications, or editorial judgment \cite{allcott2017social}. Facebook is one of the most important social media where people write their ideas and some of them who are called influencers can get a number of views as many as prominent news agencies such as Fox News and the New York Times \cite{allcott2017social}. Recent studies have shown that around 62\% of adults in the US tend to read and get the latest news on social media \cite{gottfried2016news}. On the other hand, Willmore found that the majority of false stories were being spread on Facebook more than any other mainstream social media and news stream. Many people admitted that when they encountered fake news, they believed its content \cite{willmore2016analysis}. People often use their intuition and heuristic approaches such as mental rules of thumb when they want to evaluate the authenticity of a claim and evidence. For instance, they ask themselves: 'Have I heard these stories and news before?', 'Can my knowledge help me to fit this news into my mind as a true source?'. To some extent, in many conditions, these approaches could be influential \cite{richter2009you}. One of the issues is that people do not interpret the news and misinformation objectively and in a neutral condition. However, since they have knowledge and beliefs in advance, they would like to accept the news that concurs with their point of view and beliefs \cite{johnson2009communication}. Social media platforms made it possible for fake news generators to spread their beliefs at an accelerated rate. These networks' algorithms also made it possible for users to disseminate information cheaply and quickly. So, as we go further, we see a lot of people start propagating misleading information intentionally on social media due to their political, societal, and financial motives \cite{shu2017fake}. For instance, Balmas found that Facebook was the most popular platform for the dissemination of fake news than any other mainstream news media in the 2016 US election \cite{balmas2014fake}. Aside from that, some people use these media to influence people's beliefs for their own political gain or religious purposes \cite{shu2017fake}. Their manipulation of news allows them to advance their cause by giving fake news authentication. 

Increasing literacy is one approach to confront misinformation on social media. Tully et al. tried to test this way on Twitter by enhancing people’s literacy about health issues related to genetically modified crops and the flu vaccine. Two experiments were designed to explore the effect of this approach on exposure to those fields’ misinformation and the users’ literacy about other news in general. Smith and Seitz examined the effectiveness of corrective information in debunking myths in Neuroscience. Their findings showed there was a reduction in believing the neuromyths in Facebook news feeds when readers were exposed to corrective information immediately after reading the misinformation \cite{smith2019correcting}. Van der Meer and Jin found out that corrective information had an influence on the perception of people in times of crisis such as a virus pandemic. They realized factual information caused people to take more cautionary actions than simple rebuttals. \cite{van2020seeking}. In a study, researchers delve into the rationales behind believing misinformation at an individual level. They found out that not only were some correction ways ineffective in believing misinformation but also they could backfire and make things worse. In spite of ideological beliefs and personal points of view, cognitive psychological approaches are available for rectifying misinformation beliefs, which were influential in correcting beliefs in misinformation \cite{lewandowsky2012misinformation}. Health issues always have difficulties caused by misinformation in social media. When a pandemic happens, there are rumors about the virus, hygiene and safety precautions, and vaccination. Bode and Vraga studied ways of correcting misinformation about the Zika virus on Facebook back in 2018. First, individuals with low and high conspiracy beliefs had to read misinformation. Two correction avenues were explored. A group of them accessed correct information through an algorithm and the other group accessed authentic information generated by other Facebook members. Comparing the results with a control group has shown that both ways of introducing correct information, which was through the algorithmic way and social correction, had an influence on the misperception of misinformation about the Zika virus on Facebook \cite{bode2018see}. In another study, the effect of data included in the news was explored as a way to increase the credibility of information and distinguish between information and misinformation. Five different tests were conducted by Du et al. and they reached the conclusion that when data by itself cannot increase the credibility of the news. However, when the news was presented with data visualization, people considered it a reliable source. As there was a statistical difference between news with data and news with data visualization, it could be interpreted that the appearance of presenting data increases the credibility of the news, not the data itself \cite{du2019numbers}.

Agent-based modeling (ABM) is one promising approach for studying the misinformation diffusion phenomenon. ABM allows researchers to simulate the behavior of individuals in a social network and study how misinformation spreads over time.
Tambuscio et al. \cite{tambuscio2015fact} proposed an ABM approach to study the effects of fact-checking on the spread of hoaxes in social networks. Their model found that fact-checking can significantly reduce the spread of hoaxes, especially when combined with network interventions such as blocking or suspending accounts. Sulis and Tambuscio \cite{sulis2020simulation} also used ABM to simulate misinformation-spreading processes in social networks. They demonstrated how the effects of interventions, such as limiting the number of connections or deleting nodes, can impact the spread of misinformation.
Other researchers have used ABM to explore the role of cognitive biases and social influence on the spread of misinformation. For example, Del Vicario et al. \cite{del2016spreading} used an ABM approach to study how confirmation bias can lead to the spread of false information. Their model found that individuals tend to selectively share information that confirms their pre-existing beliefs, even if it is false.
In addition to ABM, other computational approaches have been used to study the spread of misinformation, such as network analysis, natural language processing, and machine learning. For example, Lazer et al. \cite{lazer2018science} used network analysis to study the spread of false news on Twitter, while Shu et al. \cite{shu2017fake} used natural language processing to detect fake news articles. These approaches can complement ABM and provide a more comprehensive understanding of the mechanisms behind the spread of misinformation.
Research on misinformation can be focused in three directions: automatic detection of fake news \cite{conroy2015automatic}, \cite{shu2017fake}, assessing the effect and aspect of it from a psychological perspective \cite{silverman2015journalism}, \cite{lewandowsky2012misinformation}, modeling the fake news diffusion and discovering the spreading dynamics \cite{tambuscio2015fact}, \cite{budak2011limiting}, \cite{jin2013epidemiological}. 
This work belongs to the third category. Many related works in this category modeled the rumor (fake news) spreading as a virus that can infect people. In particular, these models were inspired by compartmental models in which there is a population of agents, and each agent at each point in time is in a certain state (compartment). Also, the state of agents can transit to other states with simple rules by some probability rates. The most famous models of this category are the SIR (Susceptible, Infected, and Recovered) and the SIS (Susceptible, Infected, and Susceptible). 
Different extensions of SIR-based models have been proposed with different new factors and assumptions. Among those previous works, there is a model called "SBFC" \cite{tambuscio2015fact} in which agents have three states (Susceptible, Believer, and Fact-Checker). This model describes misinformation diffusion as a competition between a piece of fake news and its debunking. Consider that we have a population of N agents, and each agent \(i\) at each time t has a state of \(S_i(t)\)  that can be one of the following:
\begin{itemize}
    \item Susceptible (S) is an agent who will change its state based on the news.
    \item Believer (B), an agent who believes in the hoax.
    \item Fact-checker (FC), an agent who has verified the news or directly knows that it is a hoax.
\end{itemize}
As it was mentioned, misinformation and fact-checking are competing actors and the main factors are the credibility of the hoax, the spreading rate, the belief of neighbors, probability of verifying or forgetting the news.

\begin{figure}
\centerline{\includegraphics{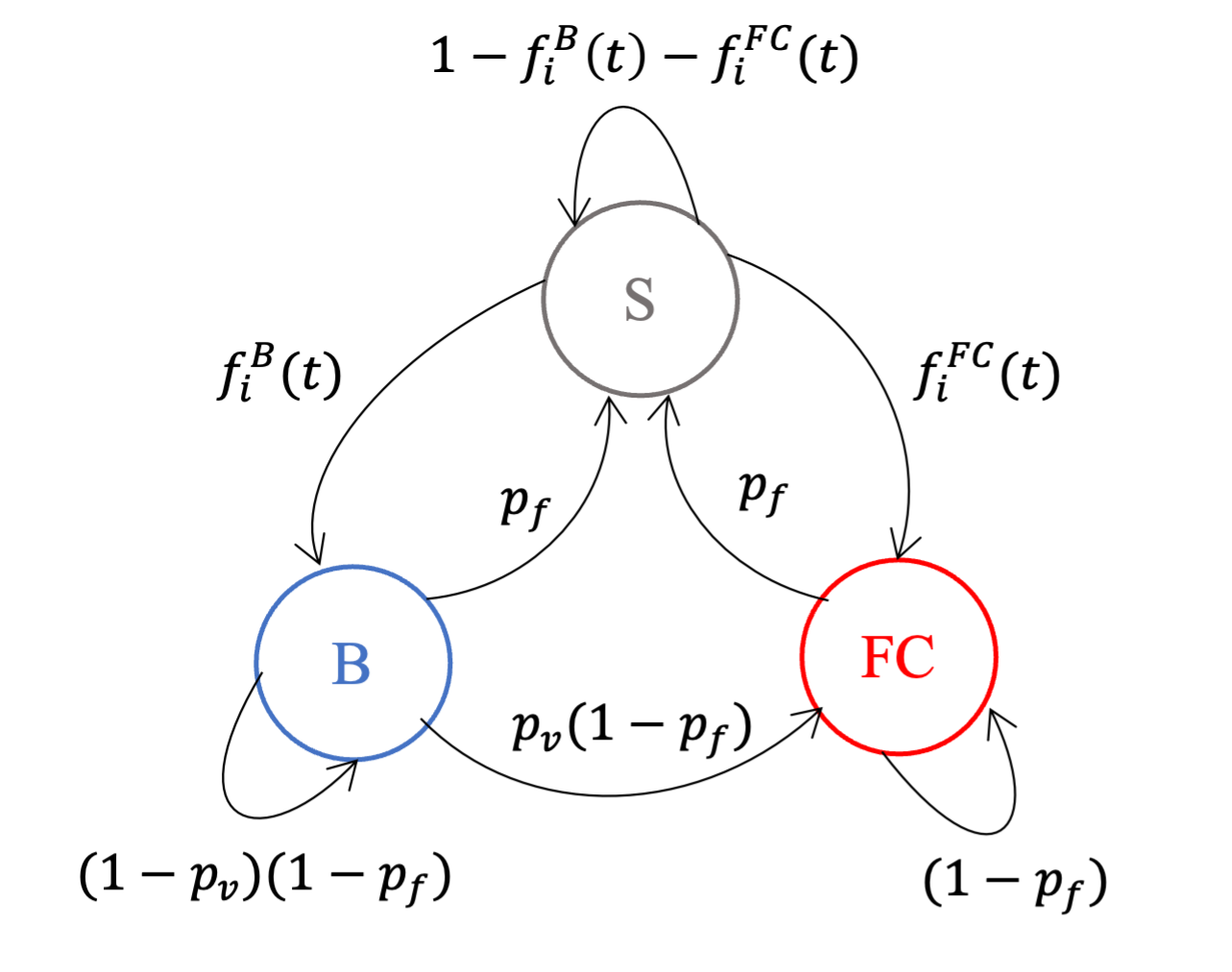}}
\caption{dynamics of SBFC model \cite{tambuscio2015fact}.}
\label{fig1}
\end{figure}

The dynamics are shown in Fig.~\ref{fig1}. There are two spreading transitions (\(f_i^B (t),f_i^{FC} (t)\)) that make a susceptible (S) agent become a believer (B) or a fact-checker (FC). These transitions are defined based on hoax credibility ($\alpha$), spreading rate ($\beta$), and belief of neighbors ($n_i^B (t)$: believers, $n_i^{FC} (t)$: fact-checkers):

\begin{equation}
f_i^B (t)= \beta \frac{n_i^B (t)(1+\alpha)}{n_i^B (t)(1+\alpha)+n_i^{FC} (t)(1-\alpha)}
\label{eq1}
\end{equation}

\begin{equation}
f_i^{FC} (t)= \beta \frac{n_i^{FC} (t)(1-\alpha)}{n_i^B (t)(1+\alpha)+n_i^{FC} (t)(1-\alpha)}
\label{eq2}
\end{equation}
\\
Also, agents who are not in the susceptible state, can forget their belief with a certain probability ($p_f$) and become susceptible again for the next time step $(t+1)$. The last factor is the probability of verifying ($p_v$). A believer agent can become a fact-checker with this probability if it does not forget the news (with the probability of $p_v (1-p_f)$).(Fig.~\ref{fig1}). 

This model was originally implemented with R programming and mean-field analysis. Sulis and Tambuscio \cite{sulis2020simulation} implemented the SBFC model in the NetLogo and got the equivalent results. They just used NetLogo for this simulation and did not extend the original work.

\section{Method}
\subsection{Social network and assumptions}
We used Netlogo to implement and test our assumptions and their effect. Our model is an extension to the SBFC model \cite{tambuscio2015fact}. So, we built the extra rules and assumptions on top of their Netlogo model \cite{sulis2020simulation}.
We have four different classes of agents in the network with their specific characteristics:
\begin{enumerate}
    \item The scholars ("Sc") who have a clearer understanding of information, can verify the news better, remember the news credibility longer, and their community is concentrated in a cluster so they get more effect from each other as one cluster.
    \item The influencers ("I") who have more influence because of their big neighborhood size, stay in their state (believer, fact-checker) longer, and could have a more verifying probability.
    \item Bots or Super-spreaders ("Bot") who do not change their state and do not forget. They just act as spreaders in the network. They can be believer bots that are programmed to spread the hoax or they can be fact-checker bots that are programmed to mitigate the misinformation diffusion.
    \item The normal ("N") agents who forget their state more frequently, and have less verifying probability.
\end{enumerate}

\begin{figure}
\centerline{\includegraphics[width=\textwidth]{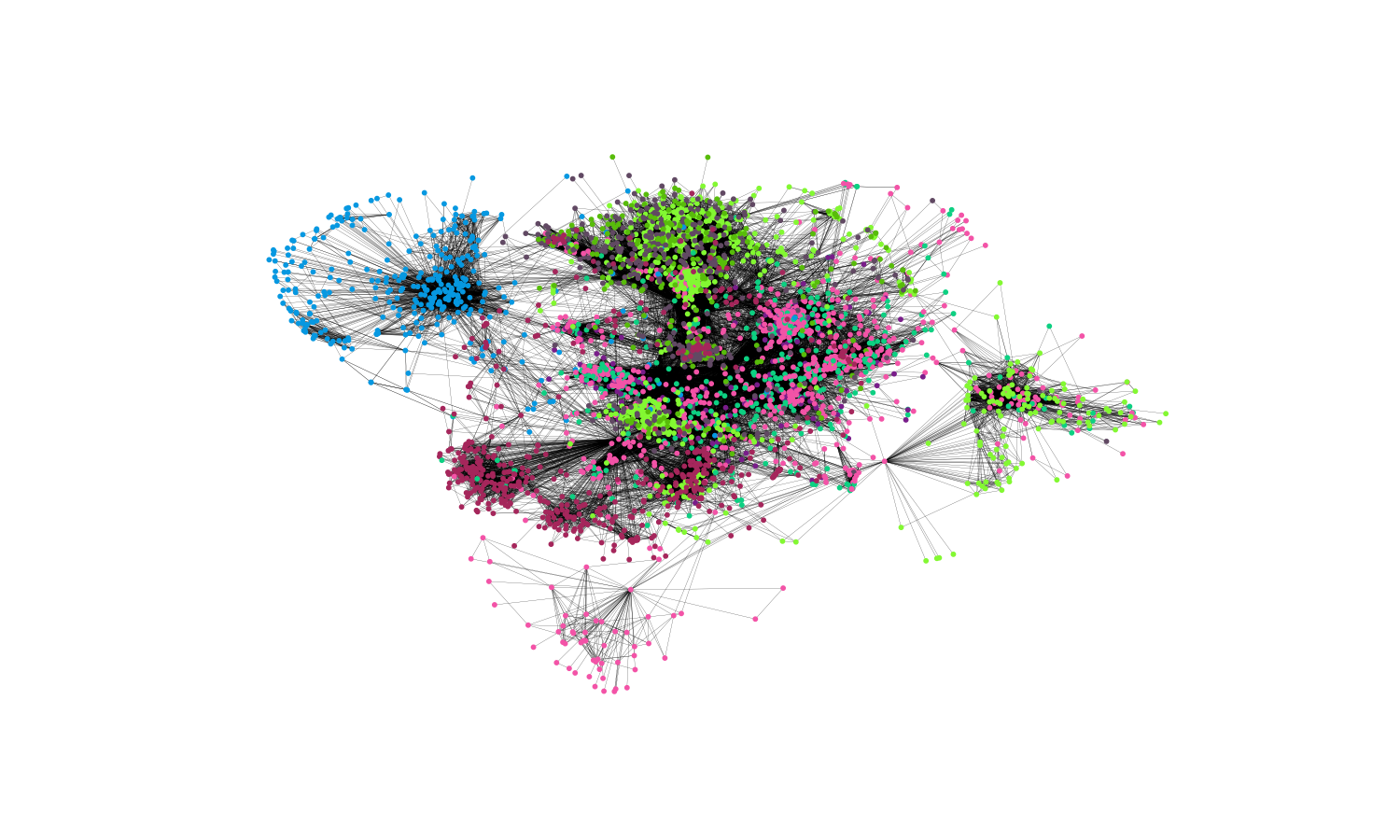}}
\caption{Facebook network with 8 clusters labeled by different colors.} 
\label{fig2}
\end{figure}

 In this work, we considered using the Facebook (social circles) dataset \cite{leskovec2012learning} instead of generating a random network in each run. This network has 4039 nodes, and 88234 edges, and the average degree of nodes is 43.7. Our goal is to analyze the behavioral change and spread of misinformation in the whole network (macro-level). With this established network, we can have fixed node degrees which are helpful for implementing the assumption of influencers. Also, we need fixed clusters because we want to assign scholars class to one of these clusters and meet our assumption of a concentrated community of this class of agents. So, having an identical network for each run is necessary for better implementing and focusing on the effect of proposed strategies.
For clustering, the asynchronous fluid communities algorithm \cite{pares2018fluid} was used to generate eight clusters (Fig. \ref{fig2}).

To implement specific characteristics, we separated the probability of verifying and forgetting the news for each class. We add initial cluster selection for scholar class which determines their percentage of the whole network. For influencers, we assume that the top 1\% of agents in terms of connectivity (node degree) are the influencers. For bots (super spreaders), we assume that the top 10\% of agents in terms of connectivity (excluding influencers) are the potential super spreaders, and bots can be assigned to them randomly with our desired percentage. Then we update the dynamics of transitions between states for each class same as the original SBFC model (Eq. \ref{eq3} and \ref{eq4}).

\begin{equation}
    n_i^B (t)= n_{{Sc}_i}^B (t) + n_{I_i}^B (t)+ n_{N_i}^B (t)+ n_{{Bot}_i}^B (t)
    \label{eq3}
\end{equation}
\begin{equation}
    n_i^F (t)= n_{{Sc}_i}^F (t)+ n_{I_i}^F (t) + n_{N_i}^F (t)+ n_{{Bot}_i}^F (t)
    \label{eq4}
\end{equation}
\\
Where the $n_i^B$ and $n_i^F$ indicate the sum of all believer and fact-checker neighbors from each class, respectively, the transition from the susceptible state to the believer or fact-checker can be calculated as before (Eq.\ref{eq1} and \ref{eq2}).

\subsection{Experimental setting}
We kept the original model \cite{sulis2020simulation} probabilities ($p_f: 0.1$, $p_v: 0.05$) for the normal (majority) class because we want to have only the changes caused by our assumptions for a meaningful interpretation. For the rest of the inputs, we considered different values (see Table \ref{tab1}) and used the behavior space tool of the NetLogo to run the simulations.

\begin{table}[ht]
\centering
\caption{Inputs information and tested values}\label{tab1}
\centering
\resizebox{\textwidth}{!}{
\begin{tabular}{ccc}
\textbf{The input name}          & \textbf{Description}                                                                                                                                                              & \textbf{values in experiments}                                                                                   \\ \hline
\textit{alpha-hoaxCredibility}   & \begin{tabular}[c]{@{}c@{}}How believable is the hoax {[}0,1{]}\\ , bigger means more believable\end{tabular}                                                                     & \{0.3, 0.8\}                                                                                                     \\ \hline
\textit{beta-spreadingRate}      & \begin{tabular}[c]{@{}c@{}}The intensity of neighbors' effect\\  on each other for spreading their belief\end{tabular}                                                            & \{0.5, 0.75\}                                                                                                    \\ \hline
\textit{\%-of-initial-believers} & \begin{tabular}[c]{@{}c@{}}The percentage of agents how are in believer state\\  at the beginning of the experiment\end{tabular}                                                  & \{10, 40\}                                                                                                       \\ \hline
\textit{scholars-community}      & \begin{tabular}[c]{@{}c@{}}The cluster of the network is assigned to be the scholar community. \\ We selected three clusters that have significant differences in size.\end{tabular} & \begin{tabular}[c]{@{}c@{}}\{"None",  Cluster4: 8.81\%, \\ Cluster7: 13.2\%, \\ Cluster8: 22.13\%\}\end{tabular} \\ \hline
\textit{pVerify-scholar}         & Probability of verifying the hoax for the scholar class                                                                                                                           & \{0.05, 0.1, 0.2, 0.3\}                                                                                          \\ \hline
\textit{pForget-scholar}         & \begin{tabular}[c]{@{}c@{}}Probability of forgetting the state and\\  become susceptible for the scholar class\end{tabular}                                                       & \{0.02, 0.05, 0.1\}                                                                                              \\ \hline
\textit{pVerify-influencer}      & Probability of verifying the hoax for the influencer class                                                                                                                        & \{0.05, 0.1, 0.2\}                                                                                               \\ \hline
\textit{pForget-influencer}      & \begin{tabular}[c]{@{}c@{}}Probability of forgetting the state and\\  become susceptible for the influencer class\end{tabular}                                                    & \{0.02, 0.05, 0.1\}                                                                                              \\ \hline
\textit{\%-of-B-bot}             & The percentage of agents how are believer bots                                                                                                                                    & \{0, 1\}                                                                                                         \\ \hline
\textit{\%-of-F-bot}             & The percentage of agents how are fact\_checker bots                                                                                                                               & \{0, 1\}                                                                                                         \\ \hline
\textit{Ticks/ Steps}            & \begin{tabular}[c]{@{}c@{}}We consider each tick as an hour and a 7-day life-cycle for news \cite{vuelioLongDoes}.\\  168 (7*24) is the total ticks/steps in each run.\end{tabular}                   & \{168\}                                                                                                          \\ \hline
\end{tabular}}
\end{table}

Since our model has stochastic nature, we generated 4 replicates for each setting. So, the 13824 different setting combinations led to 55296 total runs.

\section{Result}
\subsection{Analysis of replicates}
\begin{figure}
\centerline{\includegraphics[width=\textwidth]{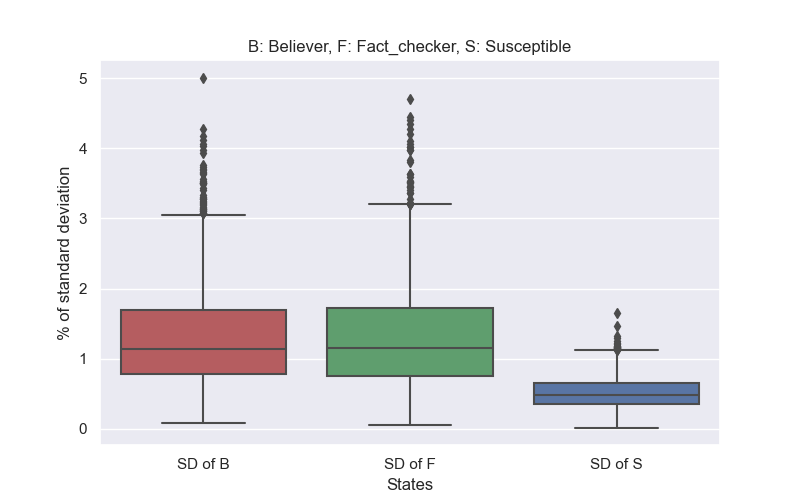}}
\caption{Standard deviation for each state, based on the replicates.}
\label{fig3}
\end{figure}

We have 4 replicates for each setting. First, we checked the result variation. To visualize this, we calculated the standard deviation of the first 2000 settings based on their replicates. Figure \ref{fig3} shows that the percentage of deviation to the total number of agents (4039) does not exceed 5\% (mostly below 2\%). So, the difference is not significant. Based on this result, we used the mean of replicates to represent each setting for the rest of the analyses. This result shows we achieved our goal of controlling the randomness at the network level, and we can genuinely explore the effects of different settings for various strategies.

\subsection{Analysis of information spread dynamics variables}
\begin{figure}
\centerline{\includegraphics[width=\textwidth]{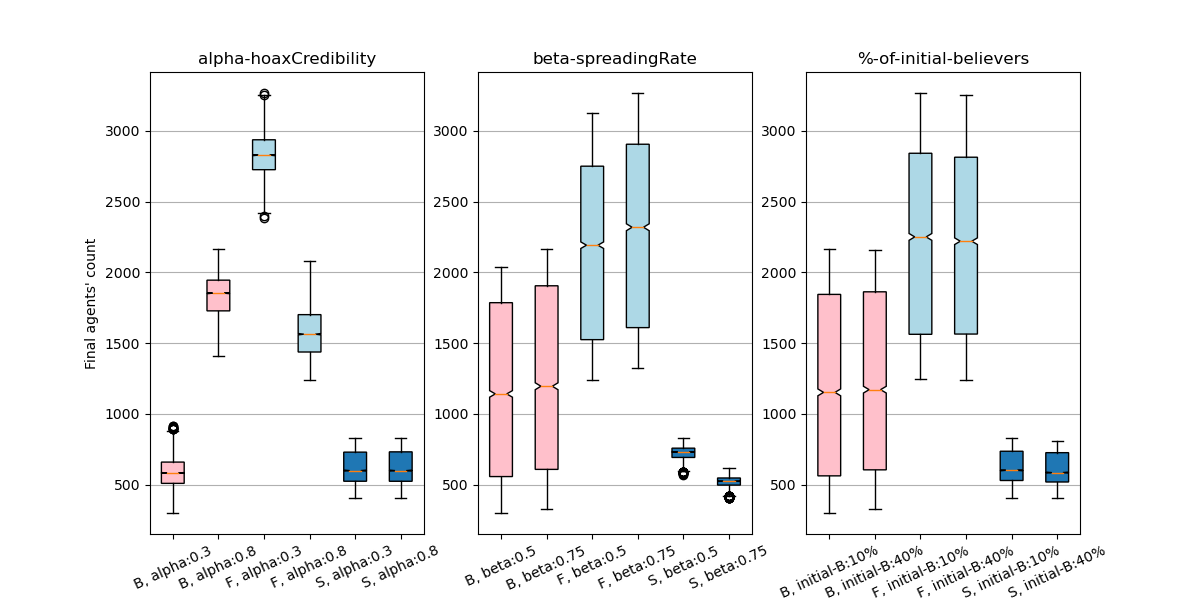}}
\caption{Difference between each input value related to spreading dynamics on final agents' count for each state. The first two columns from left are for Believers (B), the next two are for Fact-checkers (F), and the last two are for Susceptibles (S)}
\label{fig4}
\end{figure}

There are three input variables related to how the misinformation will spread. The hoax credibility ($\alpha$), the spreading rate ($\beta$), and the percentage of initial believers when the simulation starts. According to Table \ref{tab1}, we have two different values for each of these inputs. Figure \ref{fig4} shows the effect of these values and the sensitivity of each input variable. The percentage of initial believers does not show a significant difference when the value changed from 10\% to 40\%. The minor change is that when it starts with 40\% believers, the final result will contain a bit more believers and fewer fact-checkers. A higher rate of spreading ($\beta$) will decrease the final susceptible agents and turn them into either believers or fact-checkers. So it can have a positive or negative effect based on the network's majority state and other variables. The hoax credibility ($\alpha$) has the most significant sensitivity on the outcome and final states. When $\alpha = 0.3$, the fact-checker agents are the majority of the network at the end and far more than believers. With $\alpha = 0.8$, the situation is reversed and the believers' count will increase up to 4 times more than before. In this situation, intense competition will occur between believers and fact-checkers. So, further strategies and analyses will be applied to studying the solutions on how to have fewer believers in the hard circumstance ($\alpha = 0.8$).

\subsection{Analysis of new classes}
Among the newly defined classes, two classes of scholars and influencers have their specific probabilities of verifying and forgetting. The first question would be how these specified probabilities can affect the whole network. Table \ref{tab2} shows the percentage of each state at the end of the simulation based on different probabilities for each class. As mentioned, our goal is to mitigate the misinformation spread. So, we are looking for fewer believers and more fact-checkers in the final output. We define "educating" as an action of making the forgetting probability decrease and the verifying probability increase. Educating influencers will reduce the number of believers and increase the number of fact-checkers by 2\%. Educating the scholars' agents will have a greater impact (8-9\%). Although we chose scholar community number 7 (13.2\% of the total) and it is almost 13 times bigger than the influencers population (1\%), the effect intensity is not 13 times bigger. It is worth mentioning that scholars are more or less a close community and the spreading effect caused by individuals for influencers' class is more significant. Based on these conclusions, educating both classes is important. It is worth mentioning we are considering a stochastic level of cooperation after educating the agents because we only manipulated the probabilities and it is still highly possible for them to act in the opposite direction of their education goal.

\begin{table}
\centering
\caption{Comparing the effect of different forgetting and verifying probabilities for scholar and influencer classes. ($\alpha: 0.8$, $\beta: 0.5$, 10\% initial believer, scholar community: 7 (13.2\% of the total), and no bots)}\label{tab2}
\begin{tabular}{c|cccccccclclccclcl}
                              & \multicolumn{17}{c}{Scholar}                                                                                                                                                                                                                                                                                                                                                                                                                                                                                                                                                                                                                                                                                                                                                     \\ \hline
                              & \multicolumn{1}{c|}{}                                                                              & \multicolumn{2}{c|}{\begin{tabular}[c]{@{}c@{}}pf: 0.1\\ pv: 0.05\end{tabular}} & \multicolumn{2}{c|}{\begin{tabular}[c]{@{}c@{}}pf: 0.05\\ pv: 0.05\end{tabular}} & \multicolumn{2}{c|}{\begin{tabular}[c]{@{}c@{}}pf: 0.1\\ pv: 0.1\end{tabular}} & \multicolumn{2}{c|}{\begin{tabular}[c]{@{}c@{}}pf: 0.05\\ pv: 0.1\end{tabular}} & \multicolumn{2}{c|}{\begin{tabular}[c]{@{}c@{}}pf: 0.1\\ pv: 0.2\end{tabular}} & \multicolumn{2}{c|}{\begin{tabular}[c]{@{}c@{}}pf: 0.05\\ pv: 0.2\end{tabular}} & \multicolumn{2}{c|}{\begin{tabular}[c]{@{}c@{}}pf: 0.1\\ pv: 0.3\end{tabular}} & \multicolumn{2}{c|}{\begin{tabular}[c]{@{}c@{}}pf: 0.05\\ pv: 0.3\end{tabular}} \\ \cline{2-18} 
                              & \multicolumn{1}{c|}{}                                                                              & {\color[HTML]{656565} S}               & \multicolumn{1}{c|}{18\%}            & {\color[HTML]{656565} S}               & \multicolumn{1}{c|}{17\%}               & {\color[HTML]{656565} S}              & \multicolumn{1}{c|}{19\%}              & {\color[HTML]{656565} S}               & \multicolumn{1}{l|}{17\%}              & {\color[HTML]{656565} S}              & \multicolumn{1}{l|}{19\%}              & {\color[HTML]{656565} S}               & \multicolumn{1}{c|}{17\%}              & {\color[HTML]{656565} S}              & \multicolumn{1}{l|}{18\%}              & {\color[HTML]{656565} S}               & \multicolumn{1}{l|}{17\%}              \\
                              & \multicolumn{1}{c|}{}                                                                              & {\color[HTML]{CB0000} B}               & \multicolumn{1}{c|}{50\%}              & {\color[HTML]{CB0000} B}               & \multicolumn{1}{c|}{48\%}               & {\color[HTML]{CB0000} B}              & \multicolumn{1}{c|}{46\%}              & {\color[HTML]{CB0000} B}               & \multicolumn{1}{l|}{46\%}              & {\color[HTML]{CB0000} B}              & \multicolumn{1}{l|}{44\%}              & {\color[HTML]{CB0000} B}               & \multicolumn{1}{c|}{42\%}              & {\color[HTML]{CB0000} B}              & \multicolumn{1}{l|}{43\%}              & {\color[HTML]{CB0000} B}               & \multicolumn{1}{l|}{43\%}              \\
                              & \multicolumn{1}{c|}{\multirow{-3}{*}{\begin{tabular}[c]{@{}c@{}}pf: 0.1\\ pv: 0.05\end{tabular}}}  & {\color[HTML]{3531FF} F}               & \multicolumn{1}{c|}{32\%}              & {\color[HTML]{3531FF} F}               & \multicolumn{1}{c|}{35\%}               & {\color[HTML]{3531FF} F}              & \multicolumn{1}{c|}{35\%}              & {\color[HTML]{3531FF} F}               & \multicolumn{1}{l|}{37\%}              & {\color[HTML]{3531FF} F}              & \multicolumn{1}{l|}{37\%}              & {\color[HTML]{3531FF} F}               & \multicolumn{1}{c|}{40\%}              & {\color[HTML]{3531FF} F}              & \multicolumn{1}{l|}{39\%}              & {\color[HTML]{3531FF} F}               & \multicolumn{1}{l|}{40\%}              \\ \cline{2-18} 
                              & \multicolumn{1}{c|}{}                                                                              & {\color[HTML]{656565} S}               & \multicolumn{1}{c|}{18\%}              & {\color[HTML]{656565} S}               & \multicolumn{1}{c|}{17\%}               & {\color[HTML]{656565} S}              & \multicolumn{1}{c|}{18\%}              & {\color[HTML]{656565} S}               & \multicolumn{1}{l|}{17\%}              & {\color[HTML]{656565} S}              & \multicolumn{1}{l|}{18\%}              & {\color[HTML]{656565} S}               & \multicolumn{1}{c|}{17\%}              & {\color[HTML]{656565} S}              & \multicolumn{1}{l|}{18\%}              & {\color[HTML]{656565} S}               & \multicolumn{1}{l|}{17\%}              \\
                              & \multicolumn{1}{c|}{}                                                                              & {\color[HTML]{CB0000} B}               & \multicolumn{1}{c|}{49\%}              & {\color[HTML]{CB0000} B}               & \multicolumn{1}{c|}{46\%}               & {\color[HTML]{CB0000} B}              & \multicolumn{1}{c|}{47\%}              & {\color[HTML]{CB0000} B}               & \multicolumn{1}{l|}{45\%}              & {\color[HTML]{CB0000} B}              & \multicolumn{1}{l|}{44\%}              & {\color[HTML]{CB0000} B}               & \multicolumn{1}{c|}{42\%}              & {\color[HTML]{CB0000} B}              & \multicolumn{1}{l|}{43\%}              & {\color[HTML]{CB0000} B}               & \multicolumn{1}{l|}{42\%}              \\
                              & \multicolumn{1}{c|}{\multirow{-3}{*}{\begin{tabular}[c]{@{}c@{}}pf: 0.05\\ pv: 0.05\end{tabular}}} & {\color[HTML]{3531FF} F}               & \multicolumn{1}{c|}{33\%}              & {\color[HTML]{3531FF} F}               & \multicolumn{1}{c|}{37\%}               & {\color[HTML]{3531FF} F}              & \multicolumn{1}{c|}{35\%}              & {\color[HTML]{3531FF} F}               & \multicolumn{1}{l|}{38\%}              & {\color[HTML]{3531FF} F}              & \multicolumn{1}{l|}{37\%}              & {\color[HTML]{3531FF} F}               & \multicolumn{1}{c|}{41\%}              & {\color[HTML]{3531FF} F}              & \multicolumn{1}{l|}{39\%}              & {\color[HTML]{3531FF} F}               & \multicolumn{1}{l|}{41\%}              \\ \cline{2-18} 
                              & \multicolumn{1}{c|}{}                                                                              & {\color[HTML]{656565} S}               & \multicolumn{1}{c|}{18\%}              & {\color[HTML]{656565} S}               & \multicolumn{1}{c|}{17\%}               & {\color[HTML]{656565} S}              & \multicolumn{1}{c|}{18\%}              & {\color[HTML]{656565} S}               & \multicolumn{1}{l|}{17\%}              & {\color[HTML]{656565} S}              & \multicolumn{1}{l|}{18\%}              & {\color[HTML]{656565} S}               & \multicolumn{1}{c|}{17\%}              & {\color[HTML]{656565} S}              & \multicolumn{1}{l|}{19\%}              & {\color[HTML]{656565} S}               & \multicolumn{1}{l|}{17\%}              \\
                              & \multicolumn{1}{c|}{}                                                                              & {\color[HTML]{CB0000} B}               & \multicolumn{1}{c|}{48\%}              & {\color[HTML]{CB0000} B}               & \multicolumn{1}{c|}{48\%}               & {\color[HTML]{CB0000} B}              & \multicolumn{1}{c|}{47\%}              & {\color[HTML]{CB0000} B}               & \multicolumn{1}{l|}{45\%}              & {\color[HTML]{CB0000} B}              & \multicolumn{1}{l|}{43\%}              & {\color[HTML]{CB0000} B}               & \multicolumn{1}{c|}{42\%}              & {\color[HTML]{CB0000} B}              & \multicolumn{1}{l|}{42\%}              & {\color[HTML]{CB0000} B}               & \multicolumn{1}{l|}{42\%}              \\
                              & \multicolumn{1}{c|}{\multirow{-3}{*}{\begin{tabular}[c]{@{}c@{}}pf: 0.1\\ pv: 0.1\end{tabular}}}   & {\color[HTML]{3531FF} F}               & \multicolumn{1}{c|}{34\%}              & {\color[HTML]{3531FF} F}               & \multicolumn{1}{c|}{35\%}               & {\color[HTML]{3531FF} F}              & \multicolumn{1}{c|}{35\%}              & {\color[HTML]{3531FF} F}               & \multicolumn{1}{l|}{38\%}              & {\color[HTML]{3531FF} F}              & \multicolumn{1}{l|}{38\%}              & {\color[HTML]{3531FF} F}               & \multicolumn{1}{c|}{41\%}              & {\color[HTML]{3531FF} F}              & \multicolumn{1}{l|}{39\%}              & {\color[HTML]{3531FF} F}               & \multicolumn{1}{l|}{41\%}              \\ \cline{2-18} 
                              & \multicolumn{1}{c|}{}                                                                              & {\color[HTML]{656565} S}               & \multicolumn{1}{c|}{18\%}              & {\color[HTML]{656565} S}               & \multicolumn{1}{c|}{17\%}               & {\color[HTML]{656565} S}              & \multicolumn{1}{c|}{18\%}              & {\color[HTML]{656565} S}               & \multicolumn{1}{l|}{17\%}              & {\color[HTML]{656565} S}              & \multicolumn{1}{l|}{18\%}              & {\color[HTML]{656565} S}               & \multicolumn{1}{c|}{17\%}              & {\color[HTML]{656565} S}              & \multicolumn{1}{l|}{18\%}              & {\color[HTML]{656565} S}               & \multicolumn{1}{l|}{17\%}              \\
                              & \multicolumn{1}{c|}{}                                                                              & {\color[HTML]{CB0000} B}               & \multicolumn{1}{c|}{48\%}              & {\color[HTML]{CB0000} B}               & \multicolumn{1}{c|}{46\%}               & {\color[HTML]{CB0000} B}              & \multicolumn{1}{c|}{47\%}              & {\color[HTML]{CB0000} B}               & \multicolumn{1}{l|}{43\%}              & {\color[HTML]{CB0000} B}              & \multicolumn{1}{l|}{43\%}              & {\color[HTML]{CB0000} B}               & \multicolumn{1}{c|}{42\%}              & {\color[HTML]{CB0000} B}              & \multicolumn{1}{l|}{42\%}              & {\color[HTML]{CB0000} B}               & \multicolumn{1}{l|}{42\%}              \\
                              & \multicolumn{1}{c|}{\multirow{-3}{*}{\begin{tabular}[c]{@{}c@{}}pf: 0.05\\ pv: 0.1\end{tabular}}}  & {\color[HTML]{3531FF} F}               & \multicolumn{1}{c|}{34\%}              & {\color[HTML]{3531FF} F}               & \multicolumn{1}{c|}{37\%}               & {\color[HTML]{3531FF} F}              & \multicolumn{1}{c|}{36\%}              & {\color[HTML]{3531FF} F}               & \multicolumn{1}{l|}{40\%}              & {\color[HTML]{3531FF} F}              & \multicolumn{1}{l|}{38\%}              & {\color[HTML]{3531FF} F}               & \multicolumn{1}{c|}{41\%}              & {\color[HTML]{3531FF} F}              & \multicolumn{1}{l|}{40\%}              & {\color[HTML]{3531FF} F}               & \multicolumn{1}{l|}{41\%}              \\ \cline{2-18} 
                              & \multicolumn{1}{c|}{}                                                                              & {\color[HTML]{656565} S}               & \multicolumn{1}{c|}{18\%}              & {\color[HTML]{656565} S}               & \multicolumn{1}{c|}{17\%}               & {\color[HTML]{656565} S}              & \multicolumn{1}{c|}{18\%}              & {\color[HTML]{656565} S}               & \multicolumn{1}{l|}{17\%}              & {\color[HTML]{656565} S}              & \multicolumn{1}{l|}{19\%}              & {\color[HTML]{656565} S}               & \multicolumn{1}{c|}{17\%}              & {\color[HTML]{656565} S}              & \multicolumn{1}{l|}{18\%}              & {\color[HTML]{656565} S}               & \multicolumn{1}{l|}{17\%}              \\
                              & \multicolumn{1}{c|}{}                                                                              & {\color[HTML]{CB0000} B}               & \multicolumn{1}{c|}{48\%}              & {\color[HTML]{CB0000} B}               & \multicolumn{1}{c|}{46\%}               & {\color[HTML]{CB0000} B}              & \multicolumn{1}{c|}{45\%}              & {\color[HTML]{CB0000} B}               & \multicolumn{1}{l|}{44\%}              & {\color[HTML]{CB0000} B}              & \multicolumn{1}{l|}{43\%}              & {\color[HTML]{CB0000} B}               & \multicolumn{1}{c|}{42\%}              & {\color[HTML]{CB0000} B}              & \multicolumn{1}{l|}{42\%}              & {\color[HTML]{CB0000} B}               & \multicolumn{1}{l|}{42\%}              \\
                              & \multicolumn{1}{c|}{\multirow{-3}{*}{\begin{tabular}[c]{@{}c@{}}pf: 0.1\\ pv: 0.2\end{tabular}}}   & {\color[HTML]{3531FF} F}               & \multicolumn{1}{c|}{34\%}              & {\color[HTML]{3531FF} F}               & \multicolumn{1}{c|}{37\%}               & {\color[HTML]{3531FF} F}              & \multicolumn{1}{c|}{37\%}              & {\color[HTML]{3531FF} F}               & \multicolumn{1}{l|}{38\%}              & {\color[HTML]{3531FF} F}              & \multicolumn{1}{l|}{38\%}              & {\color[HTML]{3531FF} F}               & \multicolumn{1}{c|}{41\%}              & {\color[HTML]{3531FF} F}              & \multicolumn{1}{l|}{40\%}              & {\color[HTML]{3531FF} F}               & \multicolumn{1}{l|}{41\%}              \\ \cline{2-18} 
                              & \multicolumn{1}{c|}{}                                                                              & {\color[HTML]{656565} S}               & \multicolumn{1}{c|}{18\%}              & {\color[HTML]{656565} S}               & \multicolumn{1}{c|}{17\%}               & {\color[HTML]{656565} S}              & \multicolumn{1}{c|}{18\%}              & {\color[HTML]{656565} S}               & \multicolumn{1}{l|}{17\%}              & {\color[HTML]{656565} S}              & \multicolumn{1}{l|}{18\%}              & {\color[HTML]{656565} S}               & \multicolumn{1}{c|}{17\%}              & {\color[HTML]{656565} S}              & \multicolumn{1}{l|}{18\%}              & {\color[HTML]{656565} S}               & \multicolumn{1}{l|}{17\%}              \\
                              & \multicolumn{1}{c|}{}                                                                              & {\color[HTML]{CB0000} B}               & \multicolumn{1}{c|}{48\%}              & {\color[HTML]{CB0000} B}               & \multicolumn{1}{c|}{45\%}               & {\color[HTML]{CB0000} B}              & \multicolumn{1}{c|}{46\%}              & {\color[HTML]{CB0000} B}               & \multicolumn{1}{l|}{42\%}              & {\color[HTML]{CB0000} B}              & \multicolumn{1}{l|}{42\%}              & {\color[HTML]{CB0000} B}               & \multicolumn{1}{c|}{41\%}              & {\color[HTML]{CB0000} B}              & \multicolumn{1}{l|}{42\%}              & {\color[HTML]{CB0000} B}               & \multicolumn{1}{l|}{41\%}              \\
\multirow{-19}{*}{Influencer} & \multicolumn{1}{c|}{\multirow{-3}{*}{\begin{tabular}[c]{@{}c@{}}pf: 0.05\\ pv: 0.2\end{tabular}}}  & {\color[HTML]{3531FF} F}               & \multicolumn{1}{c|}{34\%}              & {\color[HTML]{3531FF} F}               & \multicolumn{1}{c|}{38\%}               & {\color[HTML]{3531FF} F}              & \multicolumn{1}{c|}{37\%}              & {\color[HTML]{3531FF} F}               & \multicolumn{1}{l|}{41\%}              & {\color[HTML]{3531FF} F}              & \multicolumn{1}{l|}{40\%}              & {\color[HTML]{3531FF} F}               & \multicolumn{1}{c|}{42\%}              & {\color[HTML]{3531FF} F}              & \multicolumn{1}{l|}{40\%}              & {\color[HTML]{3531FF} F}               & \multicolumn{1}{l|}{42\%}              \\ \cline{2-18} 
\end{tabular}
\end{table}

\begin{figure}
\centerline{\includegraphics[width=\textwidth, height=8cm]{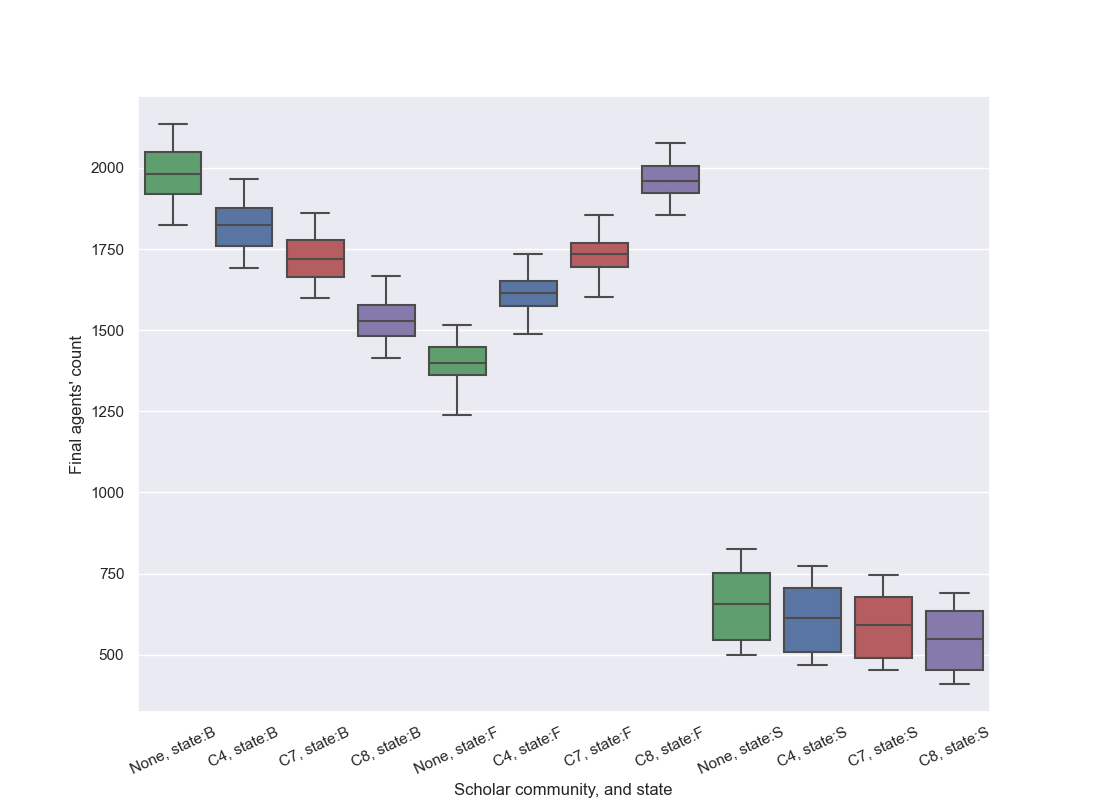}}
\caption{The effect of scholars' community size on the states of agents. ($\alpha = 0.8$, $p_{verify-scholars} = 0.3$, and $p_{forget-scholar} = 0.02$)}
\label{fig5}
\end{figure}

The next question is what would happen if we had different sizes of scholars' communities. As mentioned in table \ref{tab1}, we ran our experiments on 4 different values of scholar communities (None, 4: 8.81\%, 7: 13.2\%, 8: 22.13\%). Figure \ref{fig5} shows that a larger scholar community can be a game changer. No scholars' community (None) setting (green boxes) has a dominant population of believers and with each expansion in community size, the output will change to the favor of fact-checkers. Until community number 8 (purple boxes) has a dominant population of fact-checkers. For better discrimination, in this plot, we assumed scholars' agents are highly educated ($p_{verify-scholars} = 0.3$, and $p_{forget-scholar} = 0.02$).

\begin{figure}
\centerline{\includegraphics[width=\textwidth]{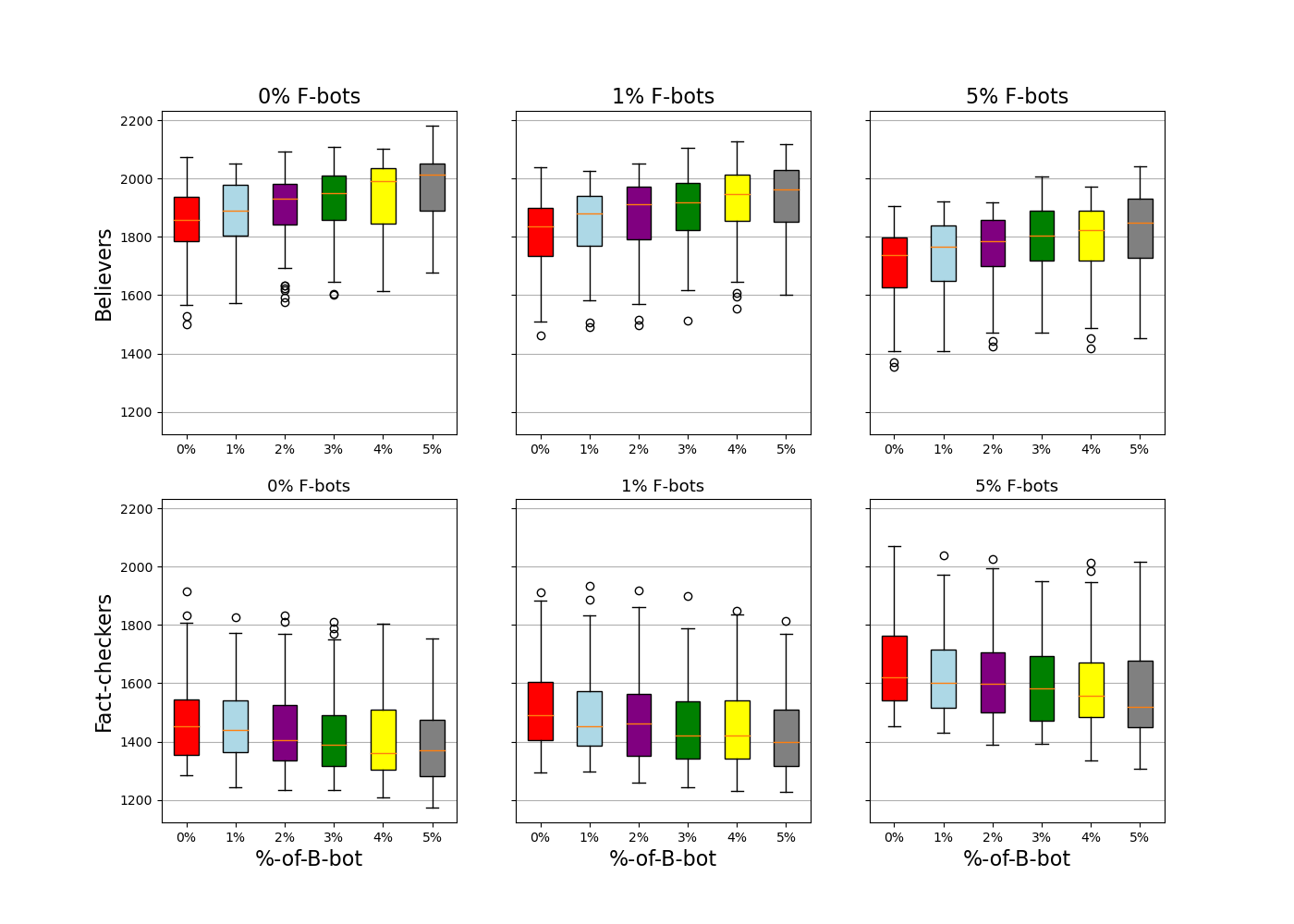}}
\caption{The effect of adding fact-checker bots for mitigating the believer bots disruption. ($\alpha = 0.8$, $\beta = 0.5$, and 10\% of initial believers)}
\label{fig6}
\end{figure}

The last class of agents in our model is the bots who are programmed to spread a certain belief. In real social media, there are some accounts that act as super spreaders and try to infect their neighbors with their beliefs. We call them bots because they do not have the ability to interpret or decide about a piece of news and do not change their state. Mostly, these agents are disruptive and try to increase the believers' agents. One possible strategy for mitigating their effect is to introduce the opposite bots (fact-checkers bots) to the social network. Since disruptive bots already exist in the network, it is ethical and even necessary for higher authorities to control their effect by any effective approaches. So, we can justify introducing the fact-checker bots only by extensive supervision of societal-level authorities. Figure \ref{fig6} shows that adding fact-checker bots can be extremely effective. Our result shows that having 5\% of each type of bots can be even better than no bots at all. Also, if we have no believer bots in the network, adding fact-checker bots can reduce the number of believers and enhance the knowledge of the network about facts. (Our initial experiment space for bots was 0 and 1\% (Table \ref{tab1}), but for more clear analysis, we added 2 - 5\% bots and conducted another experiment.)

\begin{table}[ht]
\centering
\caption{Different settings for testing the integrated effect of strategies}\label{tab3}
\begin{tabular}{cccccccc}
\multicolumn{1}{c|}{\multirow{2}{*}{\textit{\textbf{Setting}}}} & \multicolumn{3}{c|}{\textbf{Scholars}}                  & \multicolumn{2}{c|}{\textbf{Influencers}} & \multicolumn{2}{c}{\textbf{Bots}}         \\ \cline{2-8} 
\multicolumn{1}{c|}{}                                           & \textit{Community} & $p_v$ & \multicolumn{1}{c|}{$p_f$} & $p_v$     & \multicolumn{1}{c|}{$p_f$}    & \textit{Believer} & \textit{Fact-checker} \\ \hline
\multicolumn{1}{c|}{\textbf{Normal}}                            & None               & 0.05  & \multicolumn{1}{c|}{0.1}   & 0.05      & \multicolumn{1}{c|}{0.1}      & 0\%               & 0\%                   \\ \hline
\multicolumn{1}{c|}{\textbf{Worst}}                             & None               & 0.05  & \multicolumn{1}{c|}{0.1}   & 0.05      & \multicolumn{1}{c|}{0.1}      & 5\%               & 0\%                   \\ \hline
\multicolumn{1}{c|}{\textbf{Best}}                              & 8 (22\%)           & 0.2   & \multicolumn{1}{c|}{0.05}  & 0.2       & \multicolumn{1}{c|}{0.05}     & 0\%               & 5\%                   \\ \hline
\multicolumn{1}{c|}{\textbf{Moderate}}                          & 4 (9\%)            & 0.2   & \multicolumn{1}{c|}{0.05}  & 0.05      & \multicolumn{1}{c|}{0.05}     & 3\%               & 3\%                   \\ \hline
\multicolumn{8}{c}{\textit{($\alpha = 0.8$, $\beta = 0.5$, and 10\% of initial believers)}}                                                                                                                       \\ \hline
\end{tabular}
\end{table}

\subsection{Two confronting strategies}
\begin{figure}[ht]
\centerline{\includegraphics[width=\textwidth]{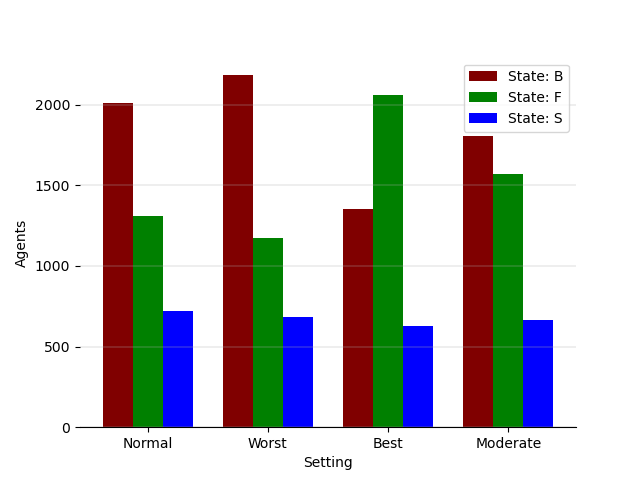}}
\caption{Change of network agents' states for different settings. Details of each setting are in Table \ref{tab3}.}
\label{fig7}
\end{figure}

Up to this point, we found out that educating minor classes of agents (scholars, or influencers) can reduce the misinformation diffusion. Another strategy is to add fact-checker bots (super spreaders) and push the crowd's belief in a positive direction. By comparing different scenarios, we can see how these two approaches can improve the states of agents. We designed four settings (see Table \ref{tab3}). Figure \ref{fig7} shows the agents' states for each setting. All of these settings are in a severe situation where the hoax credibility is too high ($\alpha = 0.8$) and it is too hard to control the misinformation spread. The 'Best' setting is employing both strategies at their full potential and the result shows that it can completely deal with the danger of misinformation because the majority of agents are in the fact-checker state. The 'Moderate' setting includes both types of bots with equal percentages and the educating strategy (first strategy) has not been applied at the maximum potential. Even with this situation, the 'Moderate' setting is much better than the 'Normal' setting.

\section{Conclusion}
In this study, we extended the 'SBFC' model by considering different classes of agents with their characteristics. This model can clearly depict the confrontation between agents who believe the false news and agents who know the facts and are fighting against the misinformation spread. That is the main reason we chose to extend this model. For future studies, other models with different dynamics can be explored. We assume that we have scholars, influencers, and two types of bots besides normal (majority) agents in the social network. These assumptions led us to propose two main strategies for dealing with misinformation diffusion. First, we can educate (train) a minor class, like scholars or influencers, to improve their ability to verify the news and remember their beliefs. The second strategy is adding fact-checker bots to the network to spread the facts and influence their neighbors' states.  Our findings show that if we employ both strategies together, the result would be very effective. Also, planning on how to get the maximum potential from agents' education programs is the key to a greater positive impact. At last, we conclude social networks with a bigger scholar community (users who have higher news verifying ability) have a better chance of dealing with misinformation spread.

\subsubsection{Availability} The Netlogo model can be found on the \href{https://www.comses.net/codebase-release/19347ba3-f7bd-4366-8f36-20eef4c8b1ef/}{CoMSES network}.

\subsubsection{Acknowledgements} This work was partially supported by the Defense Advanced Research Projects Agency (DARPA) under agreement HR00112290104 (PA-21-04-06).

%
%
%
\bibliographystyle{splncs04}
\bibliography{bibliography}
%

\end{document}